# Dirac equation with a magnetic field in 3D non-commutative phase space

LIANG Mai-Lin(梁麦林),* ZHANG Ya-Bin(张亚彬), YANG Rui-Lin(杨瑞林), ZHANG Fu-Lin(张福林)

（Physics Department, School of Science, Tianjin University，Tianjin 300072, China）

**Abstract:** For a spin-1/2 particle moving in a background magnetic field in noncommutative phase space, Dirac equation is solved when the particle is allowed to move off the plane that the magnetic field is perpendicular to. It is shown that the motion of the charged particle along the magnetic field has the effect to increase the magnetic field. In the classical limit, matrix elements of the velocity operator related to the probability give a clear physical picture: Along an effective magnetic field the mechanical momentum is conserved and the motion perpendicular to the effective magnetic field follows a round orbit. If using the velocity operator defined by the coordinate operators, the motion becomes complicated.
**Key words:** non-commutative phase space，Dirac equation，velocity operator，magnetic field
**PACS:** 02.40.Gh，03.65.-w，11.10.Nx

## 1 Introduction

To resolve the problem of infinite energies in quantum field theory, the idea of space-time non-commutativity was proposed [1]. Discoveries in string theory and M-theory that effects of noncommutative(NC) spaces may appear near the string scale and at higher energies [2-4] greatly motivate the studies in these areas. Recently, a lot of problems have been investigated on the theory of NC spaces [5-25] such as the quantum Hall effects [5-9], the harmonic oscillator [10-14], the coherent states [15], the thermodynamics [16], the classical-quantum relationship [17], the motion of the spin-1/2 particle under a uniform magnetic field [18], various kinds of relativistic oscillators [19, 20, 23, 24], etc. In [18], the particle is confined to the plane the applied magnetic field is perpendicular to. Here in this article, we discuss the case that the particle is allowed to move off the plane.

In the next section, we derive the energy spectrum and wave functions in 3D NC phase space. It is shown that the NC 3D phase space induces an effective magnetic field in a new direction. Matrix elements of velocity and momentum operators give solutions to the semiclassical equations of motion. The final section is the summary.

## 2 Three-dimensional motion

Without the loss of generality, we assume that the magnetic field is along the z-axis. In the symmetry gauge, components of the vector potential in 3D space have the form $(\hat{A}_1, \hat{A}_2, \hat{A}_3) = (-B_0\hat{x}_2/2, B_0\hat{x}_1/2, 0)$, where $B_0$ is the field strength. For a charged spin-1/2 particle moving in this background magnetic field, the Hamiltonian reads

$$\hat{H}_{3D} = c(\alpha_1\hat{p}_1 + \alpha_2\hat{p}_2 + \alpha_3\hat{p}_3) + \beta mc^2 \tag{1}$$

* E-mail addresses: mailinliang@yahoo.com.cn (Liang); mailinliang@tju.edu.cn (Liang)
The project was supported by National Natural Science Foundation of China under Grant No.11105077)

where $\hat{p}_1 = \hat{P}_1 + qB_0\hat{x}_2/2$, $\hat{p}_2 = \hat{P}_2 - qB_0\hat{x}_1/2$ and $\hat{p}_3 = \hat{P}_3$ are the mechanical momentum operators with $(\hat{P}_1, \hat{P}_2, \hat{P}_3)$ being the canonical momentum operators. From here on, small $\hat{p}_j$ ($j = 1, 2, 3$) represents the mechanical momentum operators, while the capital $\hat{P}_j$ ($j = 1, 2, 3$) stands for the canonical momentum operators. As the component $\hat{A}_3$ of the vector potential is zero, the mechanical and canonical momentum operators along the magnetic field are the same, which means $\hat{p}_3 = \hat{P}_3$. The matrices $\alpha_j$ ($j = 1, 2, 3$) and $\beta$ in (1) have the forms

$$\alpha_j = \begin{bmatrix} 0 & \sigma_j \\ \sigma_j & 0 \end{bmatrix}, \quad \beta = \begin{bmatrix} I & 0 \\ 0 & -I \end{bmatrix} \tag{2}$$

with $\sigma_j$ being the $2\times 2$ Pauli matrices and $I$ the identity matrix. In commutative spaces, there are the commutation relations $[\hat{p}_1, \hat{p}_3] = [\hat{p}_2, \hat{p}_3] = 0$ and so the operator $\hat{p}_3$ commutes with the Hamiltonian. In another word, the momentum along the z-axis is conserved. In this case, the 3D problem is a trivial extension of the 2D system. However, in 3D NC phase space, the momentum operator $\hat{p}_3$ doesn't commute with the Hamiltonian and is thus not conserved.

In NC phase space, the coordinate and momentum operators obey the commutation relations

$$\begin{aligned}
[\hat{x}_1, \hat{x}_2] &= [\hat{x}_2, \hat{x}_3] = [\hat{x}_3, \hat{x}_1] = i\mu \\
[\hat{P}_1, \hat{P}_2] &= i\nu, [\hat{P}_2, \hat{P}_3] = [\hat{P}_3, \hat{P}_1] = i\nu_0 \\
[\hat{x}_1, \hat{P}_1] &= [\hat{x}_2, \hat{P}_2] = [\hat{x}_3, \hat{P}_3] = i\hbar
\end{aligned} \tag{3}$$

with $\mu$, $\nu$ and $\nu_0$ being the NC parameters. To show clearly the NC effects between $(\hat{P}_1, \hat{P}_2)$ and $\hat{P}_3$, the two parameters $\nu$ and $\nu_0$ are written in different symbols. However, $\nu$ and $\nu_0$ can be equal numerically. The Heisenberg equations of motion for the mechanical momentum operators are

$$\begin{aligned}
\frac{d\hat{p}_1}{dt} &= \hat{\upsilon}_2 m\omega_e - \hat{\upsilon}_3 ma_0 \\
\frac{d\hat{p}_2}{dt} &= -\hat{\upsilon}_1 m\omega_e + \hat{\upsilon}_3 ma_0 \\
\frac{d\hat{p}_3}{dt} &= \hat{\upsilon}_1 m\omega_e - \hat{\upsilon}_2 ma_0
\end{aligned} \tag{4}$$

where $\hat{\upsilon}_j = c\alpha_j$ ($j = 1, 2, 3$) and

$$a_0 = v_0/(m\hbar), \omega_e = qB_0/m + v/\hbar + q^2 B_0^2 \mu/(4\hbar) \tag{5}$$

The three equations in (4) can be rewritten in the vector form as

$$\frac{d\hat{\vec{p}}}{dt} = q\hat{\vec{v}} \times \vec{B}_\theta \tag{6}$$

where

$$\vec{B}_\theta = m\omega_0 \vec{n}_{3\theta}/q, \omega_0 = \sqrt{\omega_e^2 + 2a_0^2},$$
$$\vec{n}_{3\theta} = (\sin\theta, \sin\theta, \sqrt{2}\cos\theta)/\sqrt{2} \tag{7}$$
$$\sin\theta = \sqrt{2}a_0/\omega_0, \cos\theta = \omega_e/\omega_0$$

Clearly, Equation (6) describes a charged particle moving in a effective magnetic field $\vec{B}_\theta$. The mechanical momentum operator is $\hat{\vec{p}}$ and the velocity operator is $\hat{\vec{v}} = c\vec{\alpha}$. The unit vector $\vec{n}_{3\theta}$ is along the effective magnetic field. From (5, 7), we know that the motion along the magnetic field or the parameter $v_0$ increases the effective magnetic field through $a_0$.

Time derivative of the coordinate operator gives another velocity operator $\hat{\vec{u}} = (\hat{u}_1, \hat{u}_2, \hat{u}_3)$, the components of which are

$$\hat{u}_1(t) = \frac{d\hat{x}_1}{dt} = \frac{1}{i\hbar}[\hat{x}_1, \hat{H}] = c\alpha_1\left(1 + \frac{qB_0\mu}{2\hbar}\right)$$
$$\hat{u}_2(t) = \frac{d\hat{x}_2}{dt} = \frac{1}{i\hbar}[\hat{x}_2, \hat{H}] = c\alpha_2\left(1 + \frac{qB_0\mu}{2\hbar}\right) \tag{8}$$
$$\hat{u}_3(t) = \frac{d\hat{x}_3}{dt} = -c\alpha_1\frac{qB_0\mu}{2\hbar} - c\alpha_1\frac{qB_0\mu}{2\hbar} + c\alpha_3$$

In case of $\mu = 0$, this velocity operator $\hat{\vec{u}}$ reduces to $c\vec{\alpha}$. We see that the velocity operator defined by the time derivative of the coordinates and that from the Heisenberg Equation (6) are not the same. Next we give a further analysis about the velocity operator from the point of view of probability. Using the Dirac equation $\hat{H}\psi = i\hbar\partial\psi/\partial t$ and its Hermitian conjugate, one derives the equation

$$\frac{\partial\rho}{\partial t} + \frac{1}{i\hbar}[(\hat{\vec{P}}^*\psi^\dagger)\cdot c\vec{\alpha}\psi - \psi^\dagger c\vec{\alpha}\cdot\hat{\vec{P}}\psi] = 0 \tag{9}$$

where $\rho = \psi^\dagger\psi$, which can be considered as the probability density without ambiguity. In commutative case and in the coordinate representation, the canonical momentum operator is $\hat{\vec{P}} = -i\hbar\nabla$ and its complex conjugate is $\hat{\vec{P}}^* = i\hbar\nabla$. Now, Equation (9) becomes the law of

probability conservation

$$\frac{\partial \rho}{\partial t} + \nabla \cdot \vec{J} = 0 \tag{10}$$

where $\vec{J} = \psi^{\dagger} c \vec{\alpha} \psi$ is the probability current density. In noncommutative case, the situation becomes complicated. To see the meaning of Equation (9), we make a commutative realization of the noncommutative coordinate and momentum operators

$$\hat{x}_1 = \rho(x_1 - \sigma P_2), \hat{x}_2 = \rho(x_2 - \sigma P_3), \hat{x}_3 = \rho(x_3 - \sigma P_1) \tag{11a}$$

$$\hat{P}_1 = P_1 + \nu x_2 / \hbar, \hat{P}_2 = P_2 + \nu x_3 / \hbar, \hat{P}_3 = P_3 + \nu x_1 / \hbar \tag{11b}$$

where

$$\rho = (1 + \sqrt{1 - 4\mu\nu/\hbar^2})/2, \sigma = \mu/(\hbar\rho^2) \tag{12}$$

In deriving (11a,b), the relation $\nu_0 = \nu$ has been used for mathematical simplicity. The quantities $x_j$ and $P_j$ ($j = 1, 2, 3$) in (11a, b) are the coordinate and momentum operators in commutative spaces. Using (11b) and the realization $\vec{P} = -i\hbar\nabla$, it is found that Equation (9) can still be written as (10). The current still has the form $\vec{J} = \psi^{\dagger} c \vec{\alpha} \psi$. So, $c\vec{\alpha}$ can be considered as a velocity operator related to the probability. The velocity operator $\hat{\vec{u}}$ doesn't have this meaning. From the definition (8), we can see that the velocity operator $\hat{\vec{u}}$ changes with the potential in the Hamiltonian.

The mechanical momentum operator along the effective magnetic field is

$$\hat{p}_{3\theta} = \vec{n}_{3\theta} \cdot \hat{\vec{p}} = (1/\sqrt{2})(\hat{p}_1 + \hat{p}_2)\sin\theta + \hat{p}_3 \cos\theta \tag{13}$$

The plane perpendicular to this effective magnetic field is spanned by the two orthogonal unit vectors $\vec{n}_{1\theta} = (1, -1, 0)/\sqrt{2}$ and $\vec{n}_{2\theta} = (\cos\theta, \cos\theta, -\sqrt{2}\sin\theta)/\sqrt{2}$, which are orthogonal to the effective magnetic field. The mechanical momentum operators along these two directions are

$$\begin{aligned}\hat{p}_{1\theta} &= \vec{n}_{1\theta} \cdot \hat{\vec{p}} = (\hat{p}_1 - \hat{p}_2)/\sqrt{2} \\ \hat{p}_{2\theta} &= \vec{n}_{2\theta} \cdot \hat{\vec{p}} = \cos\theta(\hat{p}_1 + \hat{p}_2)/\sqrt{2} - \hat{p}_3 \sin\theta\end{aligned} \tag{14}$$

It is not difficult to show the following commutation relations

$$\begin{aligned}[\hat{p}_{1\theta}, \hat{p}_{2\theta}] &= \mathrm{i} m\hbar\omega_0 \\ [\hat{p}_{1\theta}, \hat{p}_{3\theta}] &= [\hat{p}_{2\theta}, \hat{p}_{3\theta}] = 0\end{aligned} \tag{15}$$

Defining $\alpha_{j\theta} = \vec{\alpha} \cdot \vec{n}_{j\theta}$ ($j = 1, 2, 3$), it is found that $\vec{\alpha} \cdot \hat{\vec{p}} = \alpha_{1\theta}\hat{p}_{1\theta} + \alpha_{2\theta}\hat{p}_{2\theta} + \alpha_{3\theta}\hat{p}_{3\theta}$.

From this result and the commutations (15), we get $[\hat{p}_{3\theta}, \hat{H}] = 0$, which means the momentum along the effective magnetic field is conserved, which is similar to the commutative case. Using $\hat{p}_{j\theta}$, Equations (4) or (6) are rewritten as

$$\frac{d\hat{p}_{1\theta}}{dt} = q\hat{\upsilon}_{2\theta}B_\theta, \frac{d\hat{p}_{2\theta}}{dt} = -q\hat{\upsilon}_{1\theta}B_\theta, \frac{d\hat{p}_{3\theta}}{dt} = 0 \tag{16}$$

where $\hat{\upsilon}_{j\theta} = \hat{\vec{\upsilon}} \cdot \vec{n}_{j\theta}$. One sees that $\hat{p}_{3\theta}$ is constant. Writing the eigenstates of $\hat{p}_{3\theta}$ as $|\eta\rangle$ or $\hat{p}_{3\theta}|\eta\rangle = \eta|\eta\rangle$, the eigenvalue $\eta$ is real as $\hat{p}_{3\theta}$ is Hermitian. As the motion along the effective magnetic field is clear, next we focus on the motion perpendicular to the effective magnetic field or we consider the case $\eta = 0$.

Writing the wave function as

$$|\psi(t)\rangle = |\psi\rangle \exp\left(-\frac{\mathrm{i}}{\hbar}Et\right), |\psi\rangle = \begin{bmatrix} |\psi_1\rangle \\ |\psi_2\rangle \end{bmatrix} \tag{17}$$

the stationary Dirac equation $\hat{H}_{3D}|\psi\rangle = E|\psi\rangle$ becomes

$$mc^2|\psi_1\rangle + \vec{\sigma}\cdot\hat{\vec{p}}c|\psi_2\rangle = E|\psi_1\rangle \tag{18}$$

$$\vec{\sigma}\cdot\hat{\vec{p}}c|\psi_1\rangle - mc^2|\psi_2\rangle = E|\psi_2\rangle \tag{19}$$

From (19), we have

$$|\psi_2\rangle = \frac{\vec{\sigma}\cdot\hat{\vec{p}}c}{E+mc^2}|\psi_1\rangle \tag{20}$$

Substituting (20) into (18), after some calculations we get

$$[\hat{\vec{p}}^2c^2 + m^2c^4 - \hbar\vec{\sigma}\cdot\vec{n}_{3\theta}\omega_0 mc^2]|\psi_1\rangle = E^2|\psi_1\rangle \tag{21}$$

To solve Equation (21), we notice that $\vec{\sigma}\cdot\vec{n}_{3\theta}$ commutes with $\hat{\vec{p}}^2$ and the function $|\psi_1\rangle$ can be written as $|\psi_1\rangle = |\phi\rangle|\lambda\rangle$ with

$$\vec{\sigma}\cdot\vec{n}_{3\theta}|\lambda\rangle = \lambda|\lambda\rangle \tag{22}$$

$$[(\hat{p}_{1\theta}^2 + \hat{p}_{2\theta}^2)c^2 + m^2c^4 - \hbar\lambda\omega_0 mc^2]\phi = E^2|\phi\rangle \tag{23}$$

It is easy to see that $(\vec{\sigma}\cdot\vec{n}_{3\theta})^2 = 1$ and so the eigenvalues $\lambda = \pm 1$. The operators in (11) can be rewritten in the form

$$\hat{p}_{1\theta} = \hat{P}_{1\theta} + qB_0 \hat{X}_{2\theta}/2$$
$$\hat{p}_{2\theta} = \hat{P}_{2\theta} - qB_0 \hat{X}_{1\theta}/2 \tag{24}$$

where

$$\hat{P}_{1\theta} = (\hat{P}_1 - \hat{P}_2)/\sqrt{2},\ \hat{X}_{1\theta} = (\hat{x}_1 - \hat{x}_2)\cos\theta/\sqrt{2}$$
$$\hat{P}_{2\theta} = (\hat{P}_1 + \hat{P}_2)\cos\theta/\sqrt{2} - \hat{P}_3 \sin\theta,\ \hat{X}_{2\theta} = (\hat{x}_1 + \hat{x}_2)/\sqrt{2} \tag{25}$$

which satisfy the following commutation relations

$$[\hat{X}_{1\theta}, \hat{P}_{2\theta}] = [\hat{X}_{2\theta}, \hat{P}_{1\theta}] = 0$$
$$[\hat{X}_{1\theta}, \hat{X}_{2\theta}] = \mathrm{i}\mu\cos\theta \equiv \mathrm{i}\mu_e$$
$$[\hat{P}_{1\theta}, \hat{P}_{2\theta}] = \mathrm{i}\nu(\cos\theta + \sqrt{2}\sin\theta) \equiv \mathrm{i}\nu_e \tag{26}$$
$$\left[\hat{X}_{1\theta}, \hat{P}_{1\theta}\right] = \left[\hat{X}_{2\theta}, \hat{P}_{2\theta}\right] = \mathrm{i}\hbar\cos\theta \equiv \mathrm{i}\hbar_e$$

Defining

$$\hat{A} = \frac{\hat{p}_{1\theta} + \mathrm{i}\hat{p}_{2\theta}}{\sqrt{2m\hbar\omega_0}},\ \hat{A}^\dagger = \frac{\hat{p}_{1\theta} - \mathrm{i}\hat{p}_{2\theta}}{\sqrt{2m\hbar\omega_0}} \tag{27}$$

Equation (23) becomes

$$[2mc^2\hbar\omega_0\ (\hat{A}^\dagger\hat{A} + 1/2) + m^2c^4 - \hbar\lambda\omega_0 mc^2]|\phi\rangle = E^2|\phi\rangle \tag{28}$$

The operators (27) obey $[\hat{A}, \hat{A}^\dagger] = 1$, which means that the eigenvalues of $\hat{A}^\dagger\hat{A}$ are integers

$$\hat{A}^\dagger\hat{A}|n\rangle_A = n|n\rangle_A \tag{29}$$

where the integer $n$ takes the values $0, 1, 2, 3, \cdots$. Using the ground state $|0\rangle_A$ defined by $\hat{A}|0\rangle_A = 0$, any state $|n\rangle_A$ can be written as $|n\rangle_A = (\hat{A}^{\dagger n})|0\rangle_A/\sqrt{n!}$. Replacing $|\phi\rangle$ by $|n\rangle_A$ in (28), we get the energy spectrum

$$E_{n\lambda} = \pm\sqrt{2mc^2\hbar\omega_0(n+1/2) + m^2c^4 - \hbar\lambda\omega_0 mc^2} \tag{30}$$

In quantum field theory, the negative energy corresponds to the antiparticle.

For the motion on the plane, the state can't be described by $|n\rangle_A$ completely. For a 2D problem, we usually need two quantum numbers to describe the states of the system. Define two new operators

$$\hat{q}_{1\theta} = \hat{P}_{1\theta} - \rho\hat{X}_{2\theta},\ \hat{q}_{2\theta} = \hat{P}_{2\theta} + \rho\hat{X}_{1\theta} \tag{31}$$

When $\rho = (\nu_e + qB_0\hbar_e/2)/(\hbar_e + qB_0\mu_e/2)$, the operators (31) commute with ($\hat{p}_{1\theta}, \hat{p}_{2\theta}$). Using the operators (28), one can construct a new Hermitian operator such as $(\hat{q}_{2\theta} - \mathrm{i}\hat{q}_{1\theta})(\hat{q}_{2\theta} + \mathrm{i}\hat{q}_{1\theta})$. We write the eigenstates of the new Hermitian operator as $|k\rangle_B$, the wave function for the motion on the plane perpendicular to the effective magnetic field is finally

$$|\psi_{n\lambda k}(t)\rangle = |n\rangle_A |k\rangle_B |\lambda\rangle \exp(-iE_{n\lambda}t/\hbar) \tag{32}$$

In the classical limit or the large quantum number limit (which is the Bohr's correspondence principle), q-numbers become c-numbers. Or, the momentum operators $\hat{p}_{1\theta}, \hat{p}_{2\theta}, \hat{p}_{3\theta}$ in (16) become classical momenta $p_{1\theta}, p_{2\theta}, p_{3\theta}$. In commutative quantum mechanics, it is known that the sum of the possible matrix elements gives solutions to the classical equations [26-29] in the classical limit. Here we show that such conclusions can be applied to the noncommutative spaces. Define the sum of the possible matrix elements

$$\upsilon_j(t) = \sum_{l=0}^{\infty} \langle \psi_{l\lambda k}(t)|c\alpha_j|\psi_{n\lambda k}(t)\rangle, \quad p_j(t) = \sum_{l=0}^{\infty} \langle \psi_{l\lambda k}(t)|\hat{p}_j|\psi_{n\lambda k}(t)\rangle \tag{33}$$

As the positive and negative energies correspond to the particle and antiparticle states respectively, we calculate the quantities for the positive energy or the particle state. In this case, the matrix elements $\langle \psi_{1l\lambda k}|\hat{p}_{1,2\theta}|\psi_{1n\lambda k}\rangle$ are nonzero only when $l = n \pm 1$ (This fact can be obtained by using the inverse form of formulas (24)). Through some lengthy calculations, we have

$$\upsilon_{1\theta}(t) = \sum_{l=0}^{\infty} \langle \psi_{l\lambda k}(t)|c\bar{\alpha}|\psi_{n\lambda k}(t)\rangle \cdot \vec{n}_{1\theta}$$

$$= \langle \psi_{1n+1,\lambda k}|\left[\frac{\sigma_{1\theta}\bar{\sigma}\cdot\bar{p}c^2}{E_{n\lambda\eta}+mc^2} + \frac{\bar{\sigma}\cdot\bar{p}\sigma_{1\theta}c^2}{E_{n+1,\lambda\eta}+mc^2}\right]|\psi_{1n\lambda k}\rangle \exp\left[\frac{i}{\hbar}(E_{n+1,\lambda}-E_{n\lambda})t\right]$$

$$+ \langle \psi_{1n-1,\lambda k}|\left[\frac{\sigma_{1\theta}\bar{\sigma}\cdot\bar{p}c^2}{E_{n\lambda\eta}+mc^2} + \frac{\bar{\sigma}\cdot\bar{p}\sigma_{1\theta}c^2}{E_{n-1,\lambda\eta}+mc^2}\right]|\psi_{1n\lambda k}\rangle \exp\left[\frac{i}{\hbar}(E_{n-1,\lambda}-E_{n\lambda})t\right]$$

$$\to \frac{\langle \psi_{1n+1,\lambda k}|2p_{1\theta}c^2|\psi_{1n\lambda k}\rangle}{E_{n\lambda}+mc^2}\exp\left[\frac{i}{\hbar}(E_{n+1,\lambda}-E_{n\lambda})t\right]$$

$$+ \langle \psi_{1n-1,\lambda k}|\frac{2p_{1\theta}c^2}{E_{n\lambda}+mc^2}|\psi_{1n\lambda k}\rangle \exp\left[\frac{i}{\hbar}(E_{n-1,\lambda}-E_{n\lambda})t\right]$$

$$\to \frac{2c^2}{E_{n\lambda}+mc^2}\{\langle \psi_{1n+1,\lambda k}|p_{1\theta}|\psi_{1n\lambda k}\rangle e^{i\Omega t} + \{\langle \psi_{1n-1,\lambda k}|p_{1\theta}|\psi_{1n\lambda k}\rangle e^{-i\Omega t}\}$$

$$\to \frac{c\sqrt{2mc^2n\hbar\omega_0}}{E_{n\lambda\eta}+mc^2}[\sqrt{n+1}\exp(i\Omega t) + \sqrt{n}\exp(-i\Omega t)]$$

$$\to \frac{2c\sqrt{2mc^2n\hbar\omega_0}}{E_c+mc^2}\cos(\Omega t) \tag{34a}$$

$$\upsilon_{2\theta}(t) = \sum_{l=0}^{\infty} \langle \psi_{l\lambda\eta k}(t)|c\bar{\alpha}|\psi_{n\lambda\eta k}(t)\rangle \cdot \vec{n}_{2\theta} \to \frac{2c\sqrt{2mc^2\hbar\omega_0}}{E_c+mc^2}\sin(\Omega t) \tag{34b}$$

In the derivations, the following relations are used. In the large quantum number limit

$$E_{n\lambda} \to \sqrt{2mc^2 n\hbar\omega_0 + m^2 c^4} = E_c \tag{35a}$$

$$\frac{E_{n+1,\lambda} - E_{n\lambda}}{\hbar} = \frac{E_{n+1,\lambda} - E_{n\lambda}}{(n+1)\hbar - n\hbar} \cong \frac{\partial E_{n\lambda}}{\partial (n\hbar)} = \frac{mc^2}{E_c}\omega_0 \equiv \Omega \tag{35b}$$

$$(E_{n-1,\lambda} - E_{n\lambda})/\hbar = -mc^2\omega_0 / E_c \equiv -\Omega \tag{35c}$$

From (34), we see that $\upsilon_{1\theta}^2 + \upsilon_{2\theta}^2$ is really a constant. The forms (34) describe a round orbit. For the velocity operators (8), $u_{1\theta}^2 + u_{2\theta}^2$ is not a constant. In another word, the velocity operators defined by the time derivative of the coordinates don't give a clear physical picture. By some calculations, we also get

$$p_{1\theta}(t) \to \frac{2E_c\sqrt{2m\hbar\omega_0}}{E_c + mc^2}\cos(\Omega t), \quad p_{2\theta}(t) \to \frac{2E_c\sqrt{2m\hbar\omega_0}}{E_c + mc^2}\sin(\Omega t) \tag{36}$$

There are the relations $p_{j\theta}(t) = E_c \upsilon_{j\theta}(t)/c^2$, which agrees with the ones in special relativity.

One can check (34, 36) are the solutions to the classical equations corresponding to (6) or (16). In relativistic quantum mechanics, the eigenvalues of the velocity operator are not the actual velocity of the particle. Quantum matrix elements in the classical limit provide the desired results.

**3　Summary**

Starting from the Dirac equation, the motion of a charged spin-1/2 particle in a background magnetic field is studied in 3D NC phase space. The motion of the particle off the plane that the magnetic field is perpendicular to tends to increase the effective magnetic field. The matrix elements of velocity operators from the probability give classical solutions. The velocity operators defined by the coordinate operators are quite different from those related to the probability current.

Finally, let us analyze the effects of the noncommutativity. From (30), there is a relation between the squared energies of the two states $n+1$ and $n$

$$\frac{E_{n+1,\lambda}^2 - E_{n\lambda}^2}{2mc^2\hbar} = \omega_0 \tag{37}$$

On the noncommutative parameters, there are actually no specific results. From the free fall in a uniform gravitational field, it is said that $\mu \leq 10^{-13}\,\mathrm{m}^2$ [30]. In [31], it is pointed out that the bounds for the noncommutative parameters are

$$\mu \leq 4\times 10^{-40}\,\mathrm{m}^2,\ \nu \leq 1.76\times 10^{-61}\,\mathrm{kg}^2\mathrm{m}^2\mathrm{s}^{-2} \tag{38}$$

The appearance of such a situation is due to the fact that the present experimental techniques are difficult to detect the values of the noncommutative parameters through a direct way. Using the values $\mu = 4\times 10^{-40}\,\mathrm{m}^2,\ \nu = 1.76\times 10^{-61}\,\mathrm{kg}^2\mathrm{m}^2\mathrm{s}^{-2}$, for the electron we have

$$\frac{E_{n+1,\lambda}^2 - E_{n\lambda}^2}{2mc^2\hbar} \approx \omega_e = \frac{qB_0}{m} + \frac{\nu}{m\hbar} + \frac{q^2 B_0^2 \mu}{4m\hbar} \quad (39)$$

$$= -(1.76 \times 10^{11} b)\text{Hz} + (2 \times 10^4 + 1.3 \times 10^{-14} b^2)\text{Hz}$$

where $b$ is a constant. The value of $b$ is equal to the magnitude of the magnetic field $B_0$ when $B_0$ is measured in unit of Tesla. In deriving (39), the electronic mass $m = 9.11 \times 10^{-31}$ kg, the electronic charge $q = -1.6 \times 10^{-19}$ C and the Planck constant $\hbar = 1.05 \times 10^{-34}$ J·s have been used. The first term on the right hand side of the equality is the frequency $qB_0/m$ when the noncommutative parameters are zero. The terms in the bracket are the modification due to the noncommutativity. Noncommutativity needs super high energies, which demands that the quantum number $n$ is large and the magnetic field is super strong. For a super strong magnetic field on the surface of neutron stars $B_0 \sim 10^9$ Tesla [32], the absolute value of the modification induced by the noncommutativity is measurable. However, it is very small compared to the first term. So, to detect the noncommutative effects directly needs extremely high precision.

**References**
1 Snyder H. Phys. Rev., 1947, **71:** 38
2 Seiberg N, Witten E. JHEP, 1999, **09:** 032
3 Douglas M R, Nekrasov N A . Rev. Mod. Phys., 2001, **73:** 977
4 Szabo R. Phys. Rep., 2003, **378:** 207
5 Duval C, Horvathy P A. J. Phys. A, 2001, **34:** 10097
6 Horvathy P A. Ann. Phys., 2002, **299:** 128
7 Dayi O F, Jellal A. J. Math. Phys., 2002, **43:** 4592
8 Dulat S, Li K. Eur. Phys. J. C, 2009, **60:** 162
9 Basu B, Ghosh S. Phys. Lett. A, 2005, **346:** 133
10 Gamboa J, Loewe M, Mendez F et al. Phys. Rev. D, 2001, **64:** 067901
11 Smailagic A, Spallmcci E. Phys. Rev. D, 2002, **65:** 107701
12 LIN B S, JING S C. Phys. Lett. A, 2008, **372:** 4880
13 Scholtz F G, Gouba L, Hafrer A et al. J. Phys. A, 2009, **42:** 175303
14 Jellal A, Schreiber M, Kinani E H E. Int. J. Mod. Phys. A, 2005, **20:** 1515
15 Geloun J B, Scholtz F G. J. Math. Phys., 2009, **50:** 043505
16 Jahan A. Braz. J. Phys., 2007, **37:** 144
17 Bemfica F S, Girotti H O. Braz. J. Phys., 2008, **38:** 227
18 YUAN Y, LI K, WANG J H et al. Chin. Phys. C, 2010, **34:** 543
19 Mirza B, Narimani R, Zare S. Commun. Theor. Phys., 2011, **55:** 405
20 LI K, WANG J H, Dulat S et al. Int. J. Theor. Phys., 2010, **49:** 134
21 Smailagic A, Spallmcci E. J. Phys. A, 2002, **35:** L363
22 ZHANG P M, Horvathy P A, Ngome J P. Phys. Lett. A, 2010, **374:** 4275
23 YANG Z H, LONG C Y, QIN S J et al. Int. J. Theor. Phys., 2010, **49:** 644

24 Santos E S, de Melo G R. Int. J. Theor. Phys., 2011, **50:** 332
25 JIANG Y, LIANG M L, ZHANG Y B. Can. J. Phys., 2011, **89:** 769
26 Greenberg W R, Klein A. Phys. Rev. Lett., 1995, **75:** 1244
27 Morehead J J. Phys. Rev. A, 1996, **53:** 1285
28 HUANG X Y. Phys. Lett. A, 1986, **115:** 310
29 JIA Y W, LIU Q H, PENG X H. Acta Physica Sinica, 2002, **51:** 201(in Chinese).
30 Castello-Branco K H C, Martius A G. J. Math. Phys., 2010, **51:** 102106
31 Bertolami O, Rosa J G, Aragao C M L D et al. Phys. Rev. D, 2005, **72:** 025010
32 Duncan R C, Thompson C. Astrophysical Journal, 1992, **392:** L9